# Dynamic radiological anthropomorphic thoracic phantom with a deformable chest wall

Maria Akiki[1,2], Fadi Rouphael[1], Mazen Moussallem[1,3,4,*]

[1]Centre Hospitalier du Nord, Zgharta, Lebanon
[2]Faculty of Sciences, Lebanese University, Hadath, Lebanon
[3]Radiation Oncology Department, Centre de Traitement Médical du nord, Zgharta, Lebanon
[4]Doctoral School of Sciences and Technology, and Faculty of Public Health, Lebanese University, Tripoli, Lebanon
*Corresponding author, mazenphm@hotmail.com

November 09, 2021

**ABSTRACT**

**Purpose:** Anthropomorphic radiological phantoms with realistic anatomy and motion modeling can be used in radiotherapy to confirm dose distributions measured during motion. In this work, a prototype phantom with a deformable chest wall was created based on a real human.
**Methods:** To design and build the phantom, a technique similar to 3-dimensional printing was utilized, which involved collecting computed tomography (CT) images of a patient as a reference, contouring the organs, projecting them onto polystyrene panels, and pouring appropriate material into them. As for the breathing mechanism, balloons attached to a ventilator and an air compressor were used to periodically generate a breathing cycle, and hence chest wall movement. A silicone tumor was also implanted on the lateral basal segment in the lower lobe of the right balloon to investigate its displacement during respiration. During silent and deep breathing, qualitative and quantitative tests were conducted, and the results were compared to research conducted on real humans.
**Results:** The CT images and the shape of the phantom matches those of the real patient. However, few organs densities can be optimized. The air pressure in the ventilator was so weak that it could only create quiet breathing phantom motion on its own. However, deep breathing phantom motion was obtained manually by only using the air compressor. As a consequence, reproducibility and repeatability studies were done only for quiet breathing and results were acceptable. When compared to the real scenario, the phantom's motion amplitudes were appropriate, except in the lateral direction and abdomen section, near the diaphragm, they were negligible. In addition, the tumor displacement was predominant in the Anterior-Posterior direction rather than the Superior-Inferior direction as it should be.
**Conclusions:** As a prototype, the work was successful, nevertheless, several improvements are required, such as optimization of the shape of the mediastinum and developing a mechanical diaphragm movement system synced with a high-pressure air pump.

# INTRODUCTION

Radiation therapy is the process of treating diseases, especially malignant tumors, by exposing the affected tissues to ionizing radiation. Prior to irradiation, the patient's absorbed dose distribution is estimated in order to maximize the dose delivered to the tumor while minimizing the dose received by nearby normal tissues [1]. This step is very important since tumor control and normal tissue impact responses for most disease sites are often steep functions of radiation dosage; that is, a little change in the dose supplied (5%) can result in a substantial change in the tissue's local response (20%) thus for optimal treatment, the radiation dose must be carefully planned and delivered [2].

Using physical anthropomorphic phantoms, which are defined as inanimate surrogates for a human body or anatomic regions of interest with the intention of mimicking the internal and the external anatomy of the human body tissue for a specific procedure or experiment, is a method for monitoring the quality of radiation therapy treatments. It serves as a platform for measuring radiation distributions in human-like anatomy. These phantoms are commonly used to validate new therapeutic techniques and in investigations where anatomical features might also be important. The majority of these phantoms are created using reference anatomy [3]. However, the goal of delivering a planned dose to a target volume while minimizing it to normal tissue is restricted by the fact that the dose distribution from the treatment plan is based on a single condition of the patient anatomy, whereas the patient's anatomy and position sometimes shift between treatment fractions (interfraction motion) or during one fraction (intrafraction motion). One of the most important intrafraction motions is the respiratory movement that contributes mainly to thoracic and abdominal geometric uncertainty. As a result of these positioning inaccuracies induced by internal anatomy motion and deformation during respiration, a considerable difference between the dose prescribed and received by a target volume may be detected [4], resulting in up to a 5% inaccuracy in the localized dosage values [5]. Therefore, the intrinsic patient motion should be managed before or during the session to counteract the impacts of motion [6]. Besides, using the Monte Carlo approach to create software dynamic anthropomorphic phantoms, has its own set of challenges including the long computation time required to get statistically acceptable results. Further, setting up a specific simulation for each medical linear accelerator (LINAC) and dose determined by the Monte Carlo approach is an advanced challenge that requires particular knowledge and a lot of time [1]. On the other hand, numerous anthropomorphic phantoms have already been constructed, and most of them have either a moving diaphragm but a fixed chest wall with a skeletal structure as in respiratory motion-enabled Probe-IQ phantom [7], or without a skeletal structure as in Wilhelm phantom [6], or a fixed system with only tumor movement like the CIRS dynamic thorax phantom [8].

The aim of this study is to design, manufacture, and evaluate an anatomically accurate, dynamic, and deformable thorax that is, computed tomography (CT) compatible. Hence, an anthropomorphic phantom with realistic respiratory motion modelling, that can be utilized to study the influence of motion on tumor localization and displacements, and as a radiotherapy quality assurance method, to confirm that the given dose distributions on the phantom match the estimated does compute by the Treatment Planning System (TPS), taking into consideration physiological patient movements during treatment. The phantom body represents an average human thorax in shape, proportion, and composition. It also contains fully articulated ribs, and more importantly, it stimulates the chest wall movement during the respiration cycle which no other created phantom has executed.

# MATERIALS AND METHODS

*Phantom characteristics*

The anthropomorphic breathing phantom presented in this paper was designed for experimental research on the impact of chest wall movement on dose distribution during radiation treatment. It is composed of inflated and malleable lungs, as well as a one-piece volume formed according to the anatomy of the mediastinum and diaphragm, all contained in a skeleton and surrounded by a tissue representing muscle, fat, skin, etc. Moreover, a tumor was implanted below the right lung. The lung's movement and thorax deformation were controlled by a ventilator. The phantom's whole structure is metal-free, allowing it to be used in CT scans without any artifacts. The construction of this phantom is divided into four parts: chest skeleton fabrication, intra-thoracic organs, and tumor development, breathing unit installation, and skin and extra-thoracic tissues fabrication and sculpture. In the sections that follow, each of these features will be discussed in greater detail.

*Chest Skeleton*

To design a chest skeleton that is as realistic as possible, a new technique was developed that is comparable to 3-dimensional (3D) printing. As a prelude to executing this procedure, CT images of an average-sized patient from the archives were obtained and the bones were contoured in different colors using a radiotherapy commercial TPS: Prowess Panther$^{TM}$ version 5.2 (Prowess, Inc, CA, USA) software. This allowed identifying twelve pairs of ribs (orange and dark purple), cartilages (light blue), sternum (dark fuchsia), clavicle (dark blue), thoracic vertebrae (red), and scapulae (light fuchsia) as seen in figure 1-a.

After this phase was completed, the TPS system calculated the overall volume of the bone, which allowed us to determine how much material was needed to construct the skeleton. Since the linear attenuation coefficient ($\mu$) is proportional to the physical density of the absorbing material [9], a material with a density equivalent to that of the bone, was needed to be found. On average, a material with a density of 1.5-1.6 g/cm$^3$ can be used according to D. R. White et al. study [10]. However, the material selected must also be strong, does not easily break, and is pourable. With these exact criteria, it was determined that the Tinopoxy S.F. Floor Coating Fine Type, Series 16800 + Hardener 70 from the Tinol Paints International Co. is an excellent fit to serve as a replacement for real bones, with a density of 1.602 kg /L [11].

To ensure that the work was as precise as possible, the CT scan images of every 0.5 cm slice were printed on A3 paper to acquire the true tissue dimensions using the Autodesk AutoCAD 2021 software. Then two extruded polystyrene panels were purchased, each measuring 5 cm thick, 1.25 m long, and 0.6 m wide. To match the printed images, each panel was sliced into 50 smaller panels measuring 0.5 cm thick, 41,6 cm long, and 50 cm wide (figure 1-b). Slicing was completed using a custom-built hot wire cutter. The cutter is a Nichrome metal alloy wire with a length of 60 cm and a resistance of 20 Ohms. When fed with a 12V DC battery source, the current that passes through the wire is 0.6 A. The high current passing through the resistive wire elevates its temperature to approximately 200 degrees Celsius resulting in extremely precise cutting.

To drew the printed CT scan images on the panels these steps were followed: *(1)* The printed images were numbered as well as the small panels, to guarantee that everything is ordered; *(2)* The first picture (image number 0) was centered in the middle of the first panel (panel number 0), and then three reference points were took using 3 pins: one in the center of the CT image's reference lines, one 10 cm to the left of the center point, and one 10 cm above the center point (figure 1-c); *(3)* In order to ensure that each panel has the same 3 reference points, the first remaining panels were set on the first panel, and the 3 pins were inserted; *(4)* After that all of the panels had the same reference, the pins were re-inserted in their proper locations and all the bones were projected in the image into the panel with a pencil (figure 1-d); *(5)* These markers was a big help to draw the bones on the panel in different colors to match the printed CT scan image (figure 1-e); *(6)* To achieve a continuous shape that accurately simulates the actual human body, a drawing on the back of all the panels was performed, except for panel number 0. After that, the projection of the bone structure was drawn on each of the panel's head and tail, a pin was used to create a common hole in each bone shape drawn on the panel (head and tail), and then a mini electric drill was utilized to soften the edges to get a form that matches the current panel number and the previous panel number and hence result in a continuous sculpture (figure 1-f).

Since many joints help the thorax expand during inspiration and reduce during expiration, some techniques were introduced to execute these joints. The articulation between the ribs and the thoracic vertebrae, the ribs cartilages and the sternum, the clavicle and the sternum, and the clavicle with the scapulae, were simulated by inserting a construction thread into a 1 cm x 3 cm rectangular portion of a bicycle inner tube (figure 1-g). For the intervertebral discs, a silicone-filled between two cut car inner tubes, shaped as the vertebrae spine were utilized (figure 1-h).

To prepare the filling process, a large polystyrene panel was selected as a base and a circle of a diameter of 1.5 cm was cut out in it, and a 1.5 cm-diameter silicone-filled hose was passed through it (figure 1-f). This serves as the spinal canal. The Neo5t silicone was chosen because it has a density of 1.01 g/ml, according to its datasheet [12], which is ideal for the spinal canal since it is filled with the cerebrospinal fluid of density equal to 1.00059 g/ml [13]. To begin with the filling process and avoid leaking and mistakes, screws were utilized to guarantee that every two corresponding panels are securely fastened. Then the Tinopoxy 16800 and the hardener 70 were mixed according to the material's technical sheet [11], with a volume ratio of hardener to a base component of 1:3.09 ~ 32 % (Tinopoxy volume x 0.32 = hardener volume).

The job was extremely precise and there was a lot to prepare before filling the bone holes. For example, since the work was done in a room with a temperature of over 23 degrees, and the pot life of the mixing liquids is about 15-20 minutes, according to the technical sheet, only 200 ml of Tinopoxy with 65 ml of hardener were mixed at time using a liquid measuring cup filling the panel holes one by one (figure 1-j), while not forgetting to insert the joints into the sculpture. After completing this process, the polyester panels were removed after a couple of days resulting in the bone structure (figure 1-k).

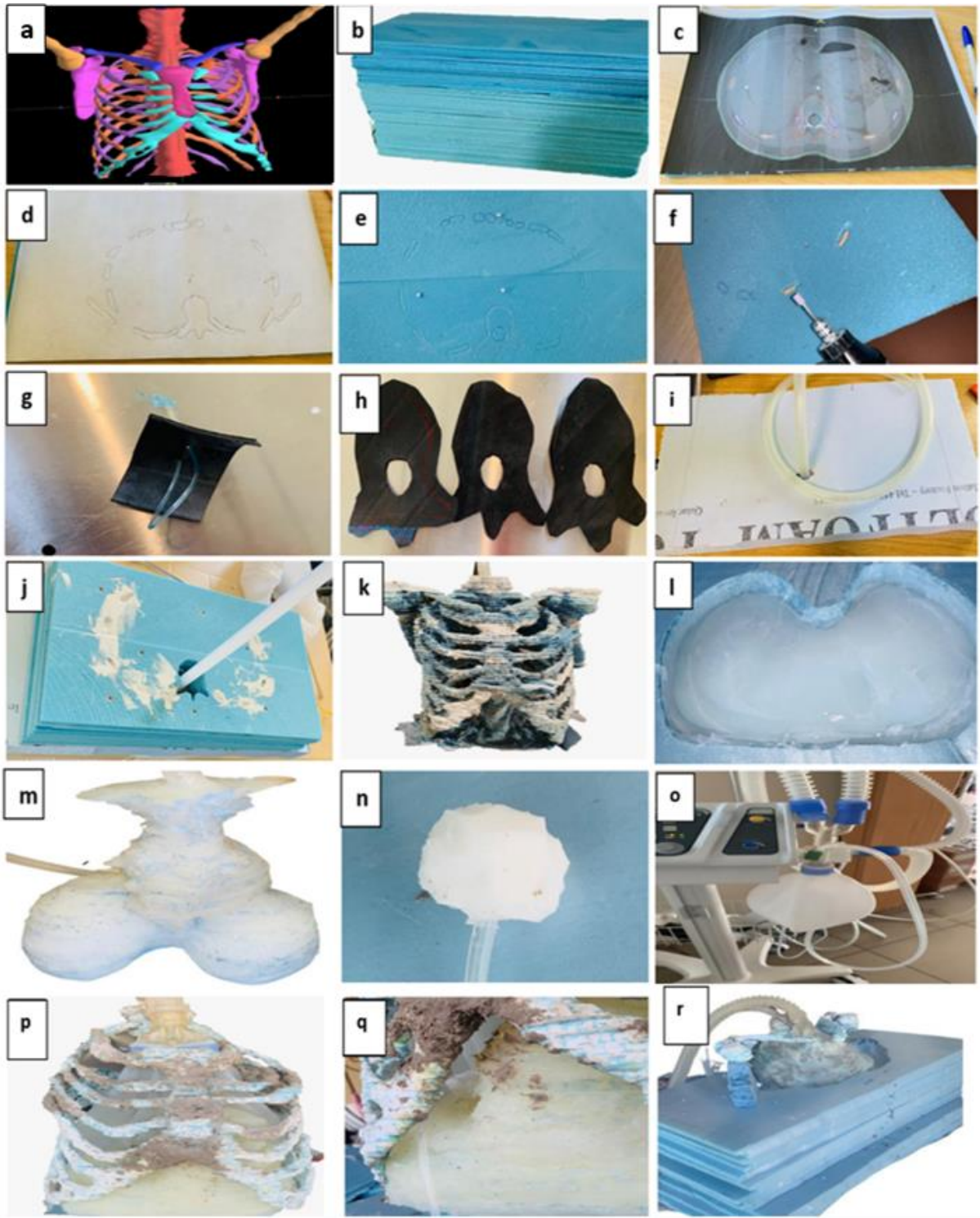

**Figure 1.** (a) The resulting contouring in 3D. (b) The resulting cut panels. (c) The 3-reference points. (d) The pencil marks on the image. (e) The drawing on the panel. (f) Electric drill tool. (g) The joint for the articulations. (h) Car inner tube-shaped as the intervertebral discs. (i) The base of the chest skeleton with the hose. (j) The filling process. (k) Final skeleton structure. (l) Silirub placed in the shaped panels. (m) The final 3D shape of the organs. (n) Tumor. (o) The balloon used along with the ventilator. (p) Organs, and balloons in the bone structure. (q) Tumor, hose location in the phantom implanted. (r) Panels embracing the skeletal structure for skin preparation.

Before moving on to the next stage, a test on the joint's articulation was done to make sure that they are working properly. Trying to articulate the newly completed structure, the amplitude of movement was fairly small. As a result, a process of decreasing the bone volumes at the joint positions was commenced by filling the gaps with silicone. After applying this to all the joints, the structure was finally articulated as it should.

### *Intra-thoracic organs and tumor*

#### *Organs*

To obtain the right morphology, all the organs including the heart, liver, and diaphragm were projected as a single 3D shape, onto the 2D panels using the same approach explained earlier but now by contouring only the organs on 1 cm x 41 cm x 30 cm polystyrene panels.

In terms of material choosing, Silirub Neo5t silicone was selected, which has a density of 1.010 $g/cm^3$, a density that is comparable to that of the heart and liver. Then it was periodically filled into the panels following the same methodology we used to fill the bone holes (figure 1-l). As for the trachea, a y-shape hose was inserted to simulate it. The final shape of the organs is shown in figure 1-m.

#### *Tumor*

The tumor was fabricated from Silastomer P25 (a two-component silicone rubber) mixed with $CACO_3$ to increase its density according to a study made by A. R Mohammad et al. [14]. This mixture helped to differentiate the tumor from the surrounding organs.

To prepare this mixture, the material-technical sheet was followed, which showed that the mixing ratio is 1:1 by volume [15]. Hence, 30 ml of component A, 30 ml of component B, and 10 g of $CACO_3$ were blended. The density of the resulting combination would be at least 1.2 g/ml. The mixture was poured into a 30 mm diameter ping pong ball, with a small hose to provide an option to introduce a dosimetry detector for reading dose distributions in the future, and after one day, when the mixture has completely cured, the ball was cut to obtain a sphere-shaped tumor. To make the shape more realistic a small portion of the sphere was removed to get the form present in figure 1-n.

### *Breathing unit*

Since the balloon and a human lung have similar features - such as being elastic, having gas-tight structures, and because its volume changes as its internal pressure changes - ventilator balloons were used to act as a replacement for real lungs and achieve the required respiratory motion. In the present study, the AEONMED Shangrila 510 s ventilator was utilized that uses the intermittent positive pressure ventilation (IPPV) operating principle (figure 1-o).

A ventilator system was made up of two pressure zones that were connected via an airway. The pressure on the device side (tube) was referred to as the airway opening pressure ($P_{ao}$), whereas the pressure on the lung (balloon in our work) side is referred to as the alveolar pressure ($P_{alv}$). The four major components of this system are: *(1)* a high-pressure gas source, *(2)* an inspiratory valve, *(3)* an expiratory valve, and *(4)* a silicone balloon. During inspiration, the inspiratory valve opens, and the expiratory valve closes and $P_{ao}$ surpasses $P_{alv}$ in value. The resulting pressure gradient forces the gas into the lungs, increasing lung volume (the balloon volume in our case). Then during expiration, the inspiratory valve closes, and the expiratory valve opens. $P_{ao}$ is now less than $P_{alv}$. The opposite pressure gradient forces gas out of the balloons, reducing balloon volume according to Boyle's law [16]. Next, several inputs were selected in order for the ventilator to work.

"Adult" was entered as the patient type, "A/C-V" as the ventilation mode, the tidal volume ($V_t$) was set to 500 ml, "Bpm" to 12 which is in the range of the normal respiratory rate for a resting adult [17], and the I:E ratio to 1:2, which means that the expiratory phase was chosen to be longer than the inspiratory phase in order to simulate normal physiologic breathing more accurately [18].

*Skin and extra-thoracic tissues*

12 small panels were cut using the same method described earlier, each measuring 3 cm thick, 60 cm long, and 30 cm wide. The method used for drawing the bones and organs on the panels was repeated, but this time for the skin contouring. Then the skin structure was pierced out was using soldering iron and then softened the edges with a mini electric drill. Silirub Neo5t was chosen, which is the same material used to form the organs because its density (1.01 g/cm$^3$) is appropriate for this task and, more importantly, it is more flexible than any other silicone in the market, with a shore A of 16 ± 5 according to its technical sheet [12]. This will help us achieve the correct respiratory movement.

The organs, the balloons (figure 1-p), and the tumor (fixed below the right lung as seen in figure 1-q) were implanted in the skeletal structure. The balloons were attached to the ventilator by a second y-shape hose. This hose was located at the skeleton in the place of the mediastinum, for this reason, a part of the mediastinum was cut in order to facilitate organs insertion in the skeleton. After that, the balloons were fairly inflated to be able to attach the exterior surface of the balloon to the interior surface of the ribs, preventing a total loss of volume and preserving the amount of air known as residual volume (RV), much like in a real-life situation. The work resumed after one day to make sure that the silicone was fully cured before proceeding to the next stage. The last stage included embracing the chest skeleton and everything inside it with the panels which were cut out for the skin and then filling everything with silicone (figure 1-r). After a couple of days of curing, the final form of the phantom was obtained (figure 2).

*Phantom imaging*

The phantom was imaged in a supine position (figure 2) utilizing a radiotherapy breast protocol in a GE Optima CT scan machine with a slice thickness of 2.5 mm. To mark 3 points, and analyze the movement of the dynamic phantom, specialized fiducial markers were attached to the phantom's surface to act as reference points in our image. Two acquisitions were done, the first was in the expiration situation when no air compressor was attached to the phantom, and the second was in deep inspiration when the balloons were filled by using an air compressor. Then the two CT series were fused by using the TPS mentioned above.

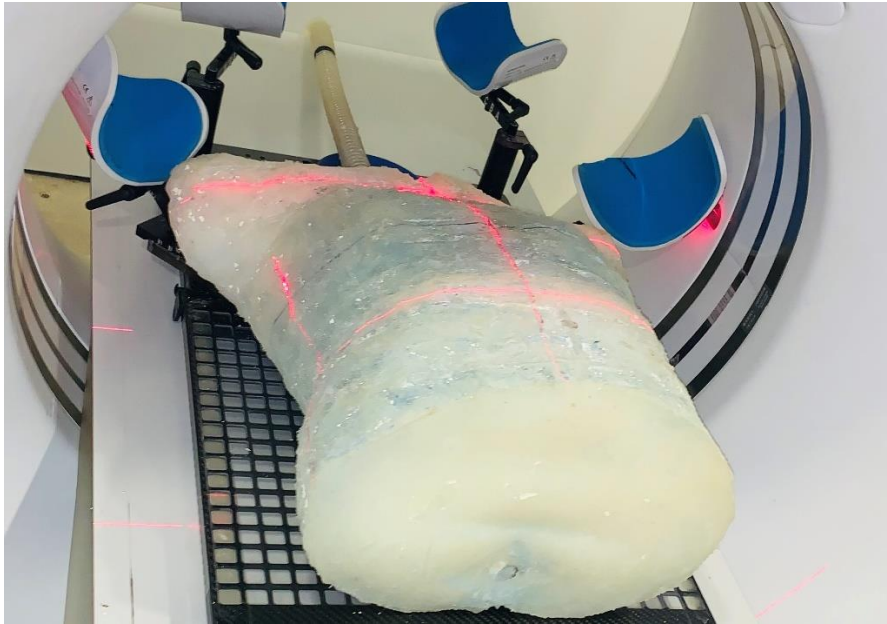
**Figure 2. Completed phantom during CT imaging.**

## *Phantom evaluation*

H. Kaneko et al. [19] used a 3D motion system (Vicon MX, Oxford Metrics, Oxford, UK) to measure the breathing movements in the supine and seated postures during quiet and deep breathing to 50 healthy males and 50 healthy females. This system was made up of eight infrared cameras that monitor the movement trajectories of 13 passive markers fixed to the thorax, chest wall, and abdominal wall. Specifically, "along the vertical line through the medial one-third of the clavicle (CL), the 3rd rib (R3), the 8th rib (R8), approximately 3 cm below the costal margin on the lateral abdomen (LAB), along the midaxillary line on the 10th rib (R10) bilaterally, and along the vertical line through the umbilicus (the sternal angle [SA], the xiphoid process [XP], the midpoint between the xiphoid process and umbilicus [abdomen - ABD])". In their paper, the locations of the 2 points on the lateral abdomen were not clearly explained, however, we deduced that they were the intersection of the vertical line of the point between the umbilicus and the xiphoid with the midaxillary line on the 10th rib. In order to compare the present phantom movement to that of real humans, these 13 points location of H. Kaneko et al. study was marked on the phantom CT images (figure 3-a) than they were projected on the phantom.

For repeatability and reproducibility study during quiet breathing, the phantom lungs were periodically inflated and deflated using a ventilator that was connected to the air compressor. Measurements were performed on each of these thirteen points using a high-precision laser meter from DEWALT. To make sure that the laser detects a constantly flat surface, a circular lapel pin was placed on every point. The whole setup is shown in figure 3-b. This meter gave us the minimum and maximum distance from a fixed rigid reference point where it is placed, and hence, it was able to determine the displacement of each of the markers at each breathing cycle. A one-minute video was taped recording the minimum and maximum readings fourteen times for each marker; seven videos for repeatability analysis and seven for reproducibility analysis, where each video contained 12 breathing cycles. The videos for repeatability analysis were taken one after the other with no change in the ventilator or phantom conditions. The reproducibility videos, however, were shot at different times, after restarting the ventilator for example and after changing the position of the phantom while keeping the laser pointed at the same marker. Furthermore, the displacement of these 13 points as well as the tumor, was detected between two CT series, one during expiration and the other during deep inspiration, and then compared to that of real human.

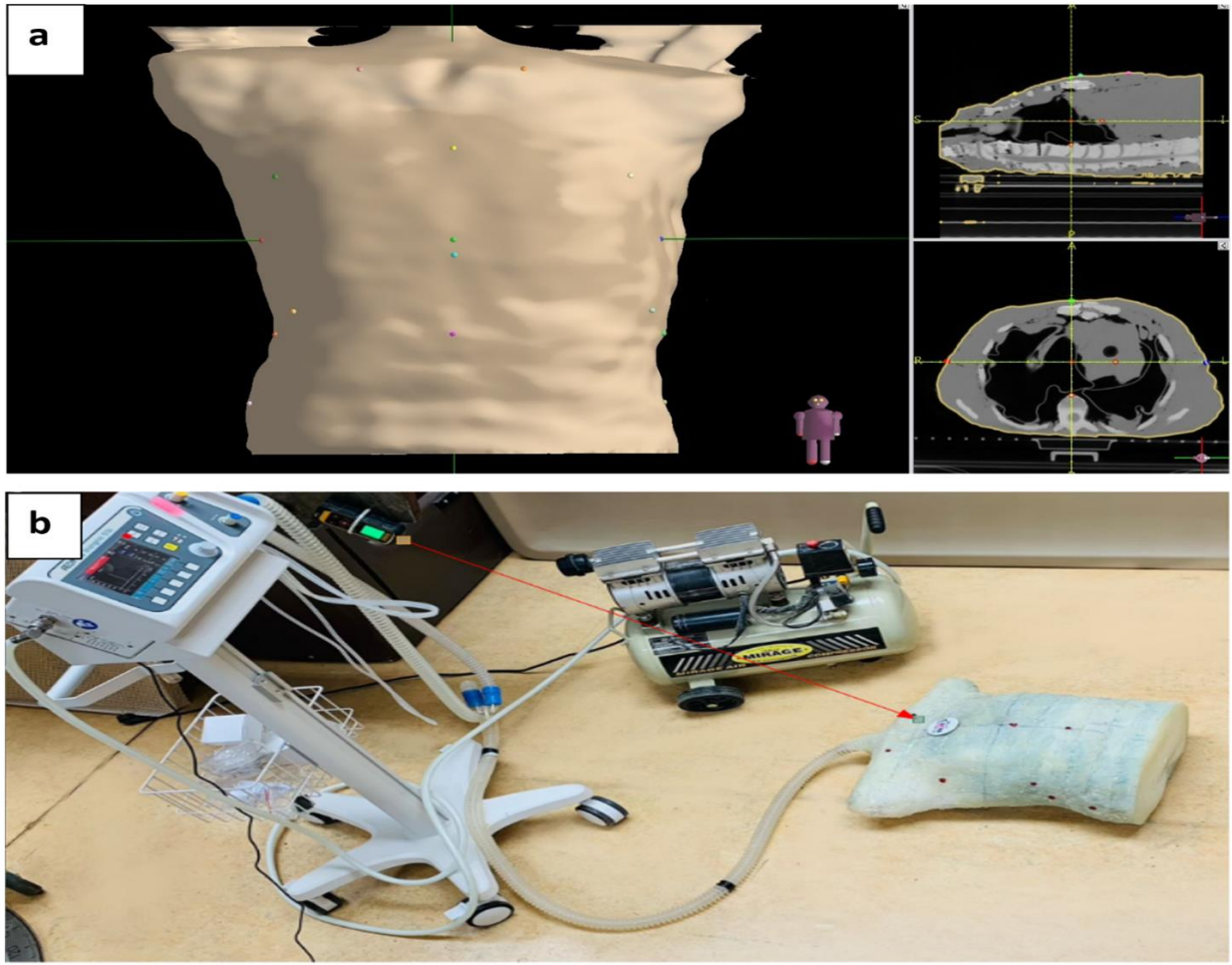

**Figure 3. (a) The thirteen points marked on the TPS. (b) The whole Set-Up.**

## RESULTS AND DISCUSSION

*Qualitative analysis*

Three images (sagittal, coronal, and axial view) were retrieved for each, the patient, and the phantom from the TPS, and were compared to one another. Figures 4-a and 4-b illustrate this. Based on these photos, it can be inferred that the phantom matches the anatomy of the patient. However, the structure of the phantom body contains air bubbles that may be improved utilizing silicone rubber (pourable liquid) instead of the Silirub Neo5t of high viscosity. Furthermore, there was a significant reduction in mediastinum size since it was cut to facilitate organ insertion into the skeleton. This issue should be considered in future work. Furthermore, the ventilator's air pressure was so low that it could only produce quiet breathing phantom motion. Deep breathing phantom motion, on the other hand, was achieved manually by only using the air compressor. As a result, reproducibility and repeatability studies were done only during quiet breathing. Thus, the usage of a high-pressure air pump ventilator must be required in future work in order to provide automated deep breathing phantom motion.

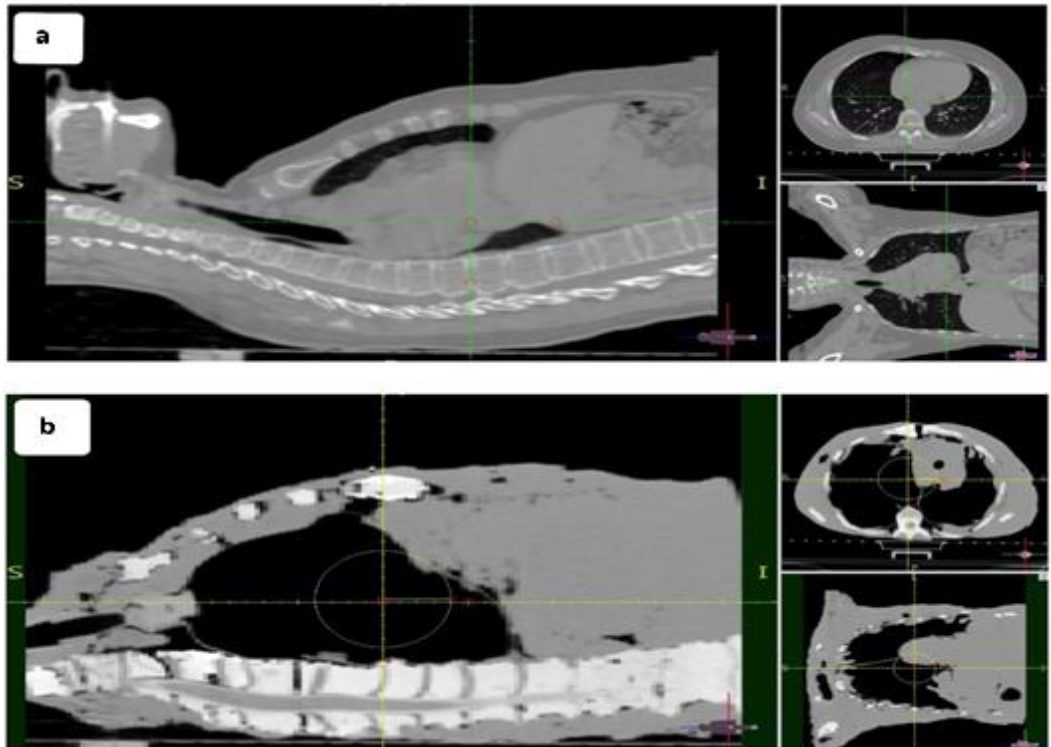

**Figure 4. (a) The real patient in sagittal view (left), axial view (upper right), coronal view (lower right). (b) The phantom sagittal view (left), axial view (upper right), coronal view (lower right).**

*Densities analysis*

The average densities result on the TPS for the phantom and the patient are shown in figure 5. This comparison shows that replacing the Tinopoxy + hardener (1.6 g/cm$^3$) with another pourable material with a density ranging from 1.048 to 1.073 g/cm$^3$, only for the cartilage part of the skeletal system, could improve our work.

To improve the density of the lungs, a foam rubber substance (also known as a sponge) might be placed inside the balloons, which has a low density due to the large number of air gaps, to over 0.085 g/cm$^3$.

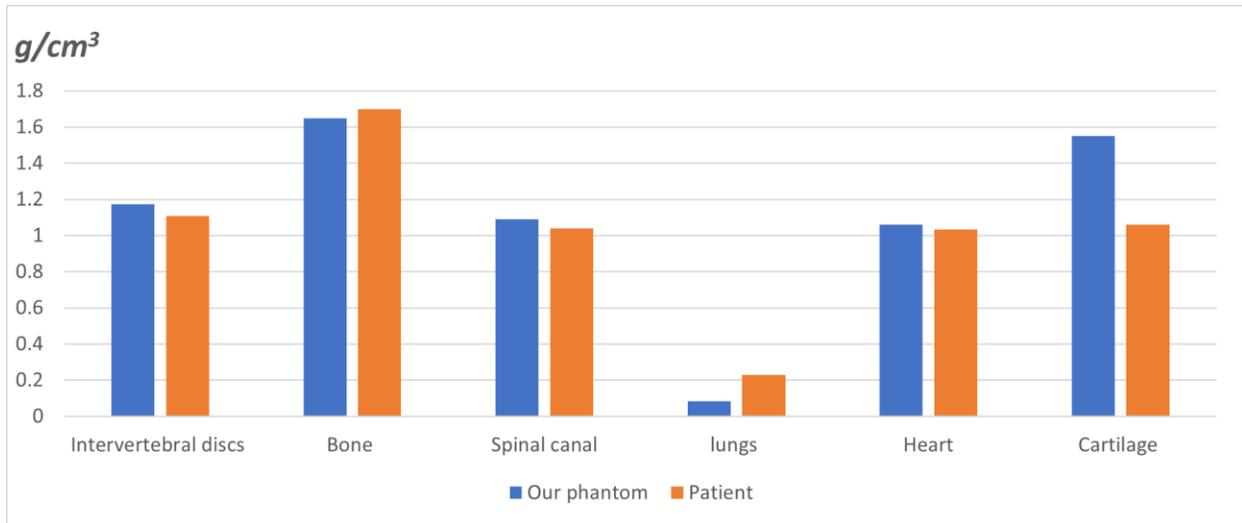

**Figure 5. Phantom vs. patient average densities.**

## *Movement analysis*

### *Quiet breathing amplitude analysis*

To ensure that the phantom's movement during passive breathing is correct, our amplitude results were compared to those of H. Kaneko et al. study [19] explained in the materials and methods section. After tabulating the repeatability results, it was revealed that every seven videos for each point have the same minimum and maximum based on a measuring accuracy of 1 mm, implying that the displacement is the same repeatedly. Therefore, the phantom is reliable and can consistently produce trustworthy outcomes. As for the reproducibility, even after the minimum and maximum measurements differed when the phantom position was changed, the difference between them, or the phantom displacement, in other words, stayed the same in each of the videos and for all the markers. The displacements were placed in figure 6 and compared to the displacements generated by H. Kaneko et al. study [19].

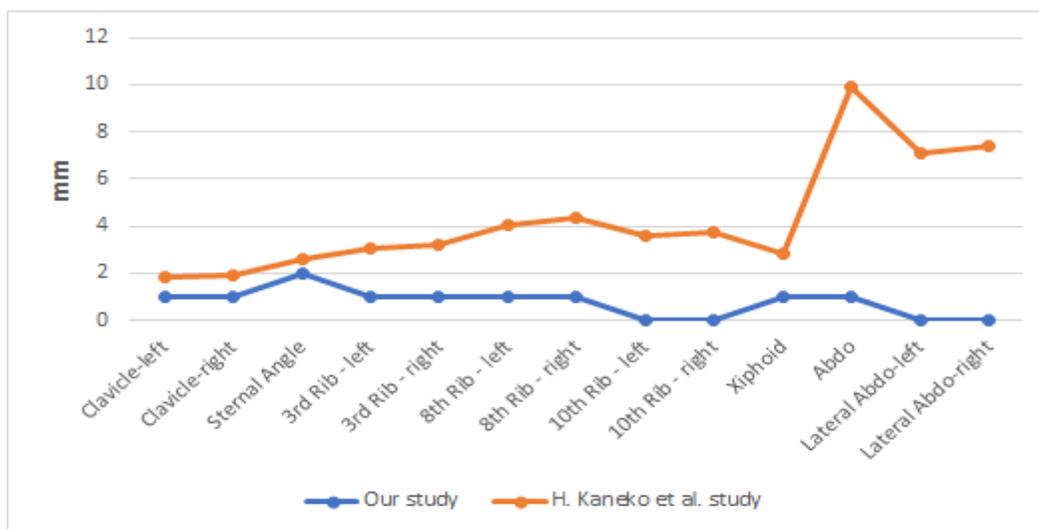

**Figure 6. Displacement comparison in mm for quiet breathing.**

As seen in the graph (figure 6), the sternal angle along with the left and right clavicle displacements are very close to those of a real human being. On the other hand, when traversed down towards the abdomen, the points started to displace in a way that is very different from a real-life situation, especially the right and left 10th rib points, the right and left lateral abdomen points, and the abdomen marker itself. The latter is logical due to the fact that the mechanism we used for breathing was lung inflation and not the true breathing mechanism which is based mainly upon diaphragm contraction. The displacement is substantial in the area where the lungs are present, but negligible in the belly area where they are not. Thus, the movement and the displacement of the phantom are highly visible in the Anterior-Posterior (A-P) direction. Furthermore, the comparison of the displacement of the 3rd rib in the two studies shows that the phantom does not move in the Medial-Lateral (M-L) direction because as mentioned above the mediastinum was cute in order to insert it into the skeleton. As a consequence, the mediastinum does not reach far enough above the 3rd rib to separate the two balloons. Consequently, when the balloons were inflated, the majority of the movement goes to fill the gap between the two balloons and not enough pressure to push the ribs outside in the lateral directions.

*Deep breathing amplitude analysis*

After the expiration and inspiration CT images were fused (figure 7) and the TPS system's features were used to measure the displacement between the 2 phases, the data was read based on the markers set on the system earlier for quiet breathing analysis in order to compare the deep breathing movement with the movement that the prior study expressed [19]. The thirteen points displacement may be seen in figure 8.

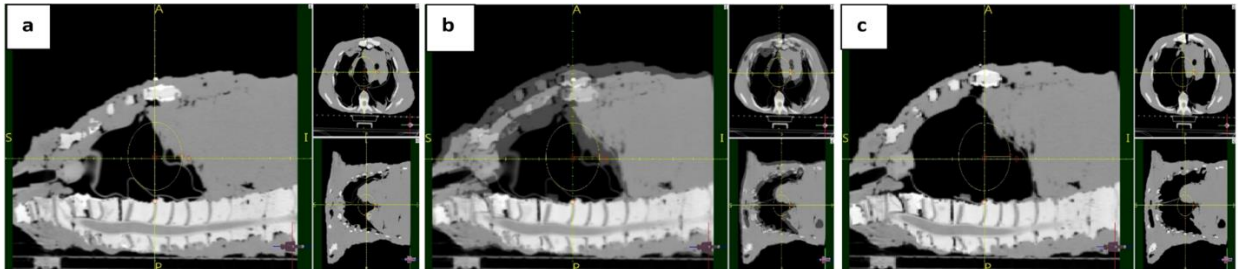

**Figure 7. Chest wall movement during inspiration and expiration. (a) Expiration. (b) Expiration and inspiration. (c) Inspiration.**

As seen in the graph, the results increased our confidence that the conclusion reached in the quiet breathing study was correct because the findings are quite similar during deep and quiet breathing. However, the displacements are more obvious during deep breathing, which is normal because the chest wall elevates more during deep breathing.

On the other hand, the tumor motion was examined using the fusion between the inspiration and expiration CT scans that were previously performed (figure 9), and the following data were obtained: 1.3 cm in the Superior-Inferior (S-I) direction, 2.4 cm in the A-P direction and 0.3 cm in the M-L direction. These results were compared to a study done by Y. Wanget al. [20]. As shown in the graph (figure 10), in real life, if a tumor is located in the right lower lobe of the lung (lateral basal segmental) and attached to the diaphragm (as in our case), the predominance of tumor motion should be in the S-I direction rather than the A-P and M-L directions. This contradicts our findings (predominance in the A-P direction) and is also due to the absence of the diaphragm movement.

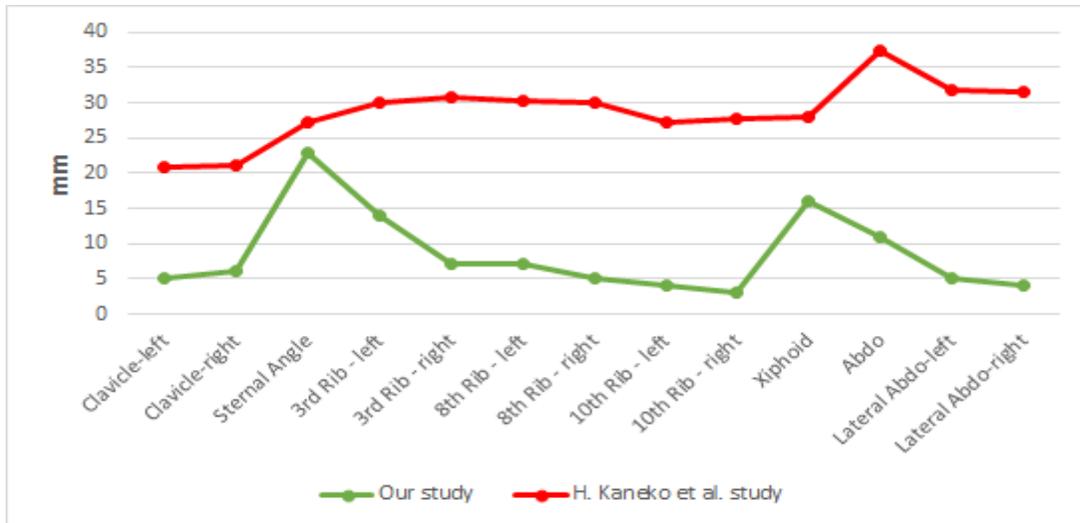

**Figure 8. Displacement comparison in mm for deep breathing.**

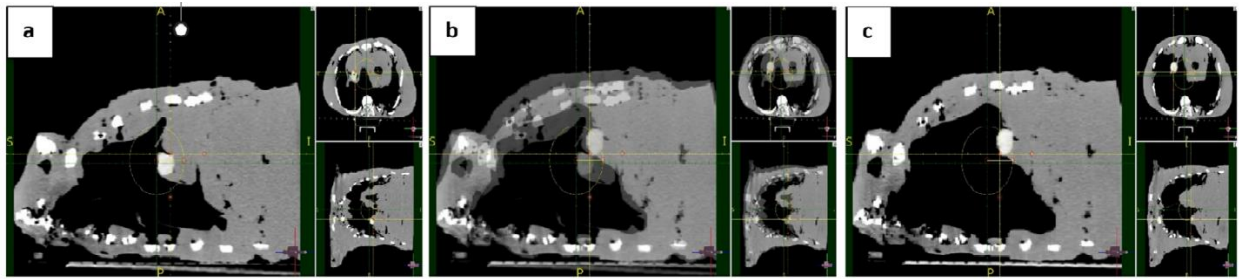

**Figure 9. Tumor movement during inspiration and expiration. (a) Expiration. (b) Expiration and inspiration. (c) Inspiration.**

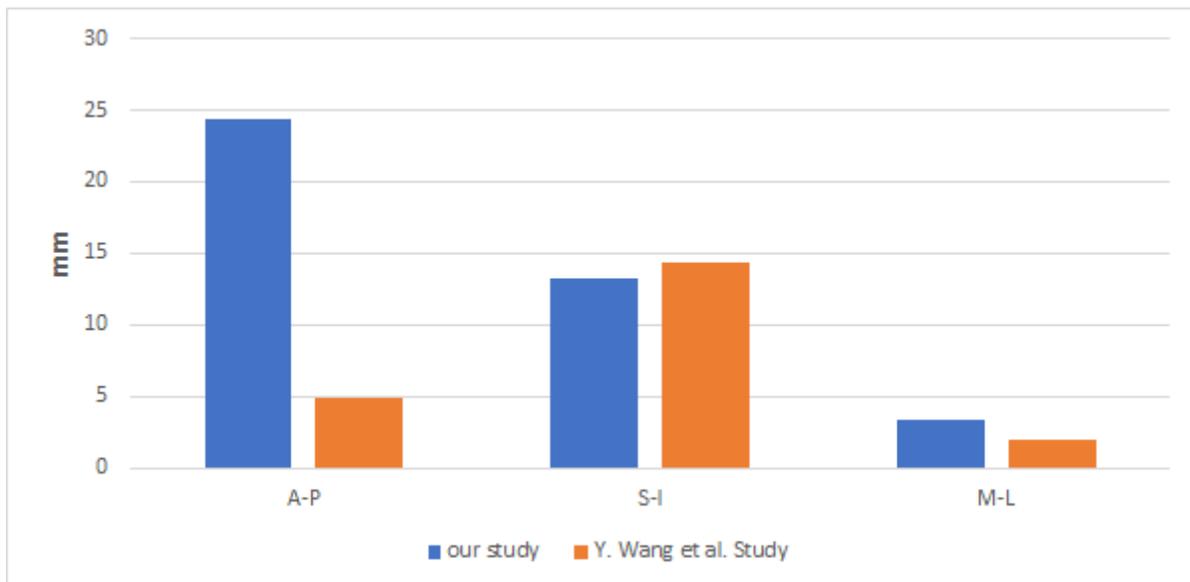

**Figure 10. Phantom vs. real human: displacement comparison of the tumor.**

Because this phantom replicates the actual anatomy and motion of a human, it, after some further optimization, makes a very promising technique that may be used to examine the influence of movement on tumor position detection. This may also be used to build, for every patient, his own phantom by following the same procedure explained in this paper. Hence, future optimizations should be done by developing a mechanical system, such as a piston and cylinder, that actuates the diaphragm in synchronous with a custom-made high-pressure air pump to produce an even better movement of the tumor and the thorax as a whole. After that, a miniature dosimetry detector such as a Thermoluminescent dosimeter (TLD) should be stuck on the skin or inserted into the tumor through the small hose implanted in it, analyze the dose distribution over the tumor and deduce the exposure of the organs at risk during the process of breathing.

## CONCLUSION

Hardware phantoms are currently either too costly or too complex to construct. In this study, a unique method for converting the structure acquired from a genuine human CT scan into a manufactured low-cost dynamic thoracic anthropomorphic phantom with chest and tumor movement was proposed. According to a thorough validation, the materials used closely reflected the relevant tissues in terms of density, shape, and mobility, and the findings were consistent, reproducible, and useful in radiation studies. However, further work will need to improve on multiple fronts, including optimizing the geometry of the mediastinum and designing a mechanical diaphragm movement system synchronized with a high-pressure air pump.

## REFERENCES

[1] Castillo, m. l. (2015). Automation of the Monte Carlo simulation of medical linear accelerators. Institut de Techniques Energètiques `.

[2] Edward, C., David, E., Carlos, A., & Luther, W. (2018). Principles and Practice of Radiation Oncology. Philadelphia : Wolters Kluwer.

[3] Halloran, A. M. (2015). Dosimetric Advantages of Personalized Phantoms. Louisiana State University and Agricultural and Mechanical College.

[4] Sara, G., Miro, V., Jason, B., Joanna, E., & Emily, H. (2017). Experimental verification of 4D Monte Carlo simulations of dose delivery to a_moving anatomy. American Association of Physicists in Medicine, 299-310.

[5] Medeiros Oliveira Ramos, S., Thomas, S., Bárbara Torres Berdeguez, M., Vasconcellos De Sá, L., & Augusto Lopes De Souza, S. (2017). Anthropomorphic Phantoms - Potential for More Studies and Training in Radiology. International Journal of Radiology & Radiation Therapy, 2(4). https://doi.org/10.15406/ijrrt.2017.02.00033

[6] Konstantin, B., Bjorn, C., Lynn, J., Florian, B., & Klaus, P. (2017). Anthropomorphic thorax phantom for cardio-respiratory motion simulation in tomographic imaging. Institute of Physiscs and Engineering in Medicine, 1-19.

[7] Black, D. G., Yazdi, Y. O., Wong, J., Fedrigo, R., Uribe, C., Kadrmas, D. J., Rahmim, A., & Klyuzhin, I. S. (2021). Design of an anthropomorphic PET phantom with elastic lungs and respiration modeling. Medical Physics, 48(8), 4205–4217. https://doi.org/10.1002/mp.14998.

[8] CIRSINC. (2013). Dynamic Thorax Phantom. https://www.cirsinc.com/wpcontent/uploads/2021/08/008A-PB-082521.pdf.

[9] Walter Huda, Richard M. Slone. Review of Radiologic Physics. (2003) ISBN : 9780781736756

[10] D. R. White, J. Booz, R. V. Griffith, J. J. Spokas, I. J. Wilson, 4. The Composition of Body Tissues, Journal of the International Commission on Radiation Units and Measurements, Volume os23, Issue 1, 15 January 1989, Pages 20–23

[11] Tinol Paints. (2019). Tinol. https://api.tinol.com/content/uploads/productsfile/4-16800.pdf.

[12] Soudal. (2020). Silirub. https://www.soudal.fr/sites/default/files/soudal_api/document/F0026681_0001.pdf.

[13] Lui, A. C. P., Polis, T. Z., & Cicutti, N. J. (1998). Densities of cerebrospinal fluid and spinal anaesthetic solutions in surgical patients at body temperature. Canadian Journal of Anaesthesia, 45(4), 297–303. https://doi.org/10.1007/bf03012018.

[14] Ahmad Rafiq Mohammad, A., Arif, F., & Dwi Cahyani, R. (2018). An Easily Made, Low-Cost, Bone Equivalent Material Used in Phantom Construction of Computed Tomography 2018 Ahmad Rafiq Mohammad Abu Arrah. Research India Publications, 7604-7609.

[15] Dalchem. (2015). http://dalchem.com.au/pdfs/Silastomer%20P25_TDS.pdf

[16] Lei, Y. (2017). Medical Ventilator System Basics : A clinical guide. Oxford University Press.

[17] Chourpiliadis, C., Bhardwaj., A., Charilaos, C., & Abhishek , B. (2021). Physiology, Respiratory Rate. Rockville Pike, Bethesda MD, 20894 USA: StatPearls Publishing.

[18] Erik, S., Devang, S., & Abhishek , B. (2021). Inverse Ratio Ventilation. 8600 Rockville Pike, Bethesda MD, 20894 USA : StatPearls Publishing LLC.

[19] Kaneko, H., & Horie, J. (2012). Breathing Movements of the Chest and Abdominal Wall in Healthy Subjects. Respiratory Care, 57(9), 1442–1451. https://doi.org/10.4187/respcare.01655.

[20] Wang, Y., Bao, Y., Zhang, L., Fan, W., He, H., Sun, Z. W., Hu, X., Huang, S. M., Chen, M., & Deng, X. W. (2013). Assessment of Respiration-Induced Motion and Its Impact on Treatment Outcome for Lung Cancer. BioMed Research International, 2013, 1–10. https://doi.org/10.1155/2013/872739.